# An F12 Correction for Multicomponent MP2

Mary J. Richardson,[1] Gabrielle B. Tucker,[1] and Kurt R. Brorsen[1]

Correspondence to Kurt R. Brorsen (Email: brorsenk@missouri.edu)

[1]Department of Chemistry, University of Missouri, Columbia, Missouri 65211, United States of America

## Abstract

Multicomponent methods treat both electrons and nuclei quantum mechanically using the standard machinery of electronic structure theory. Previous work has demonstrated that quantitatively describing electron-nuclear correlation in multicomponent many-body perturbation theory methods requires large electronic basis sets with high-angular momentum atomic orbitals, and calculations on large systems can therefore become impractical. Similarly, the slow convergence of the electron-electron correlation energy with respect to highest atomic-orbital angular momentum is a well-known issue in standard electronic structure theory in which only the electrons are treated quantum mechanically, and numerous explicitly correlated methods have been introduced to mitigate this poor convergence. Motivated by the success of single-component explicitly correlated methods, we generalize the single-component MP2-F12/3C(FIX) approach for electron-electron correlation to electron-proton correlation to reduce the need for large electronic basis sets in multicomponent calculations. We also implement the complementary auxiliary basis set (CABS) singles correction for the electrons and protons. Our calculations show that the multicomponent MP2-F12 electron-proton correlation energy is within a few percent of the complete-basis-set (CBS) estimate obtained by QZ-5Z inverse-cubic extrapolation with PB4-D held fixed at every electronic basis-set cardinality tested, with the aug-cc-pVDZ basis set recovering 96–98% of the extrapolated CBS electron-proton correlation energy and exceeding the conventional multicomponent MP2 value at the aug-cc-pV5Z level. These results indicate that explicitly correlated methods can substantially reduce the electronic basis-set requirements of multicomponent calculations.

### Acknowledgements
This material is based upon work supported by the National Science Foundation under Grant No. 2418760. We thank Prof. S. Ten-no for comments on the Gaussian representation of the Slater geminal.

### Statements and Declarations
The authors have no competing interests to disclose.

### Author Contributions
All authors contributed to the study conception and design. Data collection and analysis were performed by Mary J. Richardson and Gabrielle B. Tucker. The first draft of the manuscript was written by Kurt R. Brorsen and all authors commented on previous versions of the manuscript. All authors read and approved the final manuscript.

## I. Introduction

In electronic structure theory, the basis set truncation error of the correlation energy formally scales as $L^{-3}$, where $L$ is the maximum angular momentum atomic orbital included in the basis set.[1,2] This slow convergence is a well-known issue, and it makes quantitative calculations with highly accurate many-body methods such as coupled-cluster theory difficult because of the high polynomial scaling of these methods. Part of the reason for this slow convergence is due to the difficulty in describing the electron-electron (or Kato) cusp in the wave function[3,4] as most one-particle basis sets are constructed from Gaussian functions that cannot efficiently model this cusp.

To improve the basis set convergence of electronic structure theory methods, the past few decades have seen the introduction of geminal-based explicitly correlated methods. These methods augment the standard electronic structure theory approach by including geminals in the wave function that depend explicitly on the electron-electron separation and can therefore better model the electron-electron cusp.[5,6] Explicitly correlated methods are an extensive field of quantum chemistry, with many different methods and approximations, such that we cannot properly review the field in full here. In this study, we focus on extending the F12 methodology to treat electron-proton correlation. For a more in-depth discussion of standard F12 theory, we direct the interested reader to several F12 reviews published over a decade ago that remain an excellent introduction to the field.[7–10] In most cases, these F12 methods are able to compute energies that are at least one basis set cardinal higher than those computed by a non-F12 method (e.g. DZ→TZ) without increasing the formal asymptotic computational scaling.[11,12] Because of this, F12 methodology has been applied to many of the widely used methods of quantum chemistry including second-order Møller-Plesset perturbation theory,[6,13,14] coupled-cluster (CC) theory,[12,15,16] and multireference perturbation theory,[17–19] among many others.

One subfield of electronic structure theory in which F12 methodology has not been commonly applied is in multicomponent methods. Multicomponent methods are used to include nuclear quantum effects in computational chemistry calculations by treating electrons and other types of particles (typically select nuclei) quantum mechanically and on equal footing using generalizations of the standard or single-component methods of electronic structure theory.[20] Analogous to the electron-electron correlation energy in single-component quantum chemistry, the central quantity in multicomponent quantum chemistry is the electron-nuclear or electron-proton correlation energy. The basic ideas of multicomponent methods were introduced by Thomas in the late 1960s[21,22] and, since then, many similar frameworks have been introduced with the nuclear-electronic orbital framework likely being the most prominent.[23–28] The last decade has seen a rapid increase in the development of multicomponent methods and similar to the many different single-component F12 methods, there now exist many multicomponent generalizations of standard quantum chemistry methods including density functional theory (DFT),[29–33] many-body perturbation and coupled-cluster theory,[34–45] and configuration-interaction related methods.[28,46–50]

It is well established in the literature that large electronic basis sets are necessary for obtaining quantitative accuracy with multicomponent methods, with the basis-set truncation error dominated by the electronic basis and multicomponent CC calculations commonly requiring 5Z or larger basis sets.[51] While there have been some recent efforts to introduce electronic basis sets appropriate for multicomponent calculations, these efforts remain preliminary, and no consensus on best practices for such electronic basis sets has emerged.[50–52]

Given the need for large electronic basis sets in multicomponent calculations and the present lack of purpose-driven multicomponent one-particle electronic basis sets, we anticipate that generalizing the F12 methodology to multicomponent methods would be beneficial. Electrons and protons satisfy a generalized Kato cusp condition analogous to that of the electron-electron cusp.[4] We hypothesize that part of the need for large electronic basis sets in multicomponent calculations is due to the incomplete treatment of this cusp when using Gaussian one-particle basis sets for the electrons and protons.

Explicit correlation has been investigated in the context of multicomponent methods via the explicitly correlated Hartree-Fock approach and related methods,[53–60] but this research has largely been abandoned due to the high-polynomial scaling of these methods with respect to system size ($\geq N^8$). To our knowledge, the only use of modern F12 methodology in multicomponent methods is a study that used it to accelerate the convergence of the electron-electron correlation.[61] Modern F12 methodology has not been applied to the electron-proton correlation energy that is the central quantity of interest in most multicomponent methods.

In this study, we derive and implement an F12 correction for multicomponent MP2. We also implement the complementary auxiliary basis set (CABS) singles corrections for the electronic and protonic species. In general, we find that multicomponent MP2-F12 recovers the electron-proton correlation energy estimated by QZ–5Z extrapolation at a fixed PB4-D nuclear basis set to within a few percent at the level of the aug-cc-pVDZ electronic basis set, exceeding conventional multicomponent MP2 with the aug-cc-pV5Z electronic basis set. Based on the results of this study, we hypothesize that using the F12 approach in combination with more accurate as well as more computationally expensive multicomponent methods such as multicomponent CC could prove to be a useful direction for future research.

## II. Methodology

### II.A Multicomponent HF and MP2

In multicomponent methods, the system is partitioned into electrons, quantum nuclei, and classical nuclei. In this study, we treat selected hydrogen nuclei quantum mechanically, so we label the quantum nuclear subsystem with *p* and the classical nuclear subsystem with *N*. The multicomponent Hamiltonian is then

$$\hat{H} = \hat{T}_e + \hat{T}_p + \hat{V}_{ee} + \hat{V}_{ep} + \hat{V}_{pp} + \hat{V}_{eN} + \hat{V}_{pN} + \hat{V}_{NN}, \quad (1)$$

where $\hat{T}_{\mathrm{e}} = -\frac{1}{2}\sum_i \nabla_i^2$ and $\hat{T}_{\mathrm{p}} = -\frac{1}{2}\sum_\kappa (1/m_\kappa)\,\nabla_\kappa^2$ are the electronic and protonic kinetic energy operators, respectively, and the sums range over all electrons, indexed by *i,* or protons, indexed by $\kappa$. We assume atomic units where $m_e = 1$ and $m_p = 1836.152673$. $\hat{V}_{ee}$, $\hat{V}_{ep}$, $\hat{V}_{pp}$, $\hat{V}_{eN}$, $\hat{V}_{pN}$, *and* $\hat{V}_{NN}$ are the Coulombic electron-electron, electron-proton, proton-proton, electron-classical nuclei, proton-classical nuclei, and classical nuclei-classical nuclei Coulomb operators, respectively. The quantum proton-proton repulsion is present only for $N_q \geq 2$.

We assume that the quantum protons are distinguishable particles and can therefore neglect exchange terms among them. For most chemical systems, this approximation is reasonable as the protons are much more localized than the electrons of the system. Additionally, this approximation has been shown to introduce negligible error for multicomponent MP2[62] and is a fundamental approximation in the constrained nuclear-electronic orbital family of methods.[63] In this study, we also neglect all proton-proton correlation as a recent multicomponent coupled-cluster study showed that such correlation is nearly one-thousand times smaller than electron-proton correlation.[43]

With these assumptions, the reference multicomponent Hartree-Fock (HF) wave function is written as

$$|\Phi_{\mathrm{HF}}\rangle = |\psi^e\rangle \otimes \bigotimes_{\kappa=1}^{N_q} |\psi^{p^\kappa}\rangle, \tag{2}$$

where $|\psi^e\rangle$ is a closed-shell Slater determinant of the electronic molecular orbitals, $|\psi^{p^\kappa}\rangle$ is the molecular orbital for nucleus $\kappa$, and $N_q$ is the number of quantum nuclei.

In conventional multicomponent MP2,[34,62] the zeroth-order Hamiltonian is taken to be the sum of the electronic and protonic Fock operators. The second-order energy correction is written as a sum of electron-electron and electron-proton terms as

$$E_{\mathrm{MP2}} = E_{\mathrm{MP2}}^{ee} + \sum_{\kappa=1}^{N_q} E_{\mathrm{MP2}}^{ep^\kappa} \tag{3}$$

where $E_{\mathrm{MP2}}^{ee}$ is the conventional closed-shell electronic MP2 energy expression,

$$E_{\mathrm{MP2}}^{ee} = \sum_{ijab} \frac{\langle ij \,|\, r_{12}^{-1} \,|\, ab\rangle \left( 2 \langle ij \,|\, r_{12}^{-1} \,|\, ab\rangle - \langle ij \,|\, r_{12}^{-1} \,|\, ba\rangle \right)}{\epsilon_i^e + \epsilon_j^e - \epsilon_a^e - \epsilon_b^e}, \quad (4)$$

where $i, j$ and $a, b$ run over the occupied and virtual electronic orbitals, respectively, $r_{12}^{-1}$ is the Coulomb operator, $\langle ij \,|\, r_{12}^{-1} \,|\, ab\rangle$ is a two-electron Coulomb integral, and $\epsilon$ is the molecular-orbital energy eigenvalue. $E_{\mathrm{MP2}}^{ep^\kappa}$ is the electron-proton MP2 correlation energy of proton $\kappa$,

$$E_{\mathrm{MP2}}^{ep^\kappa} = 2 \sum_{iaIA} \frac{\left| \langle iI \,|\, r_{12}^{-1} \,|\, aA\rangle \right|^2}{\epsilon_i^e + \epsilon_I^{p^\kappa} - \epsilon_a^e - \epsilon_A^{p^\kappa}}, \quad (5)$$

where $I, A$ are the occupied and virtual protonic orbitals, respectively, of nucleus $\kappa$. The prefactor of 2 in $E_{\mathrm{MP2}}^{ep^\kappa}$ is due to the closed-shell electron-spin multiplicity. In $E_{\mathrm{MP2}}^{ep^\kappa}$, there is no exchange integral because the electron and proton are distinguishable particles.

We briefly mention that multicomponent MP2 does not always perform well for modeling nuclear quantum effects due to the multicomponent HF reference being qualitatively incorrect. In many cases, orbital optimization,[38,39] higher excitations,[41] or more advanced many-body methods such as coupled-cluster theory[36,42] are needed for quantitative accuracy, but multicomponent MP2 has begun to see some success in the constrained multicomponent formalism[44] as well as in double-hybrid electron-proton correlation functionals.[33] However, even with this caveat, we emphasize that the results of this study showing the utility of F12 methods in multicomponent MP2 should generalize to other multicomponent methods, that an F12 correction to multicomponent MP2 is much easier to implement than more advanced methods such as coupled-cluster theory and that many of the intermediates derived below can be reused in future F12 studies. Taken together, these considerations justify starting the study of interfacing F12 methodology with multicomponent methods with multicomponent MP2-F12.

### II.B Conventions

In multicomponent MP2-F12, the electrons and each of the distinguishable protons has a one-particle basis set associated with it as in conventional multicomponent MP2. We denote this basis set as the orbital basis set (OBS). To approximate completeness in the explicitly correlated intermediates, we augment each particle's OBS with a separate, larger Gaussian basis set called the auxiliary basis set (ABS). The union of the OBS and the ABS is called the resolution-of-the-identity basis set (RIBS), RIBS = OBS ∪ ABS, which is the working basis in which we insert approximate completeness relations, $\sum_p |p\rangle\langle p| \approx \mathbf{1}$, to factor F12 matrix elements into computable products. The complementary auxiliary basis set (CABS) is the orthogonal complement of the OBS within the RIBS. That is, the CABS space is the subspace of the RIBS that lies outside the span of the OBS molecular orbitals. This choice is the CABS+ construction of Valeev,[64] distinguished from the original CABS prescription by including the OBS itself in the RIBS rather than using only the ABS for the RI.[65] The specific OBS, ABS, and CABS choices for the electronic and protonic sectors are tabulated in Sections III and IV.

**II.C Multicomponent MP2-F12**

As the multicomponent MP2 energy partitions into separate electron-electron and electron-proton contributions, in this study we focus only on including the F12 correction for the electron-proton MP2 correlation energy as the existing single-component F12 methodology can be used with no changes to correct the electron-electron MP2 correlation energy. With this assumption, we augment the first-order multicomponent MP2 wave function with a single Slater geminal, $f_{12}$, that correlates the electrons to each proton. We do this using Ansatz 3 of Werner and coworkers[66] as

$$\left|\Psi^{(1)}\right\rangle = \left|\Psi^{(1)}_{\mathrm{OBS}}(t)\right\rangle + \sum_{\kappa=1}^{N_q} \sum_{iI} t_{iI}^{(\kappa)}\, \hat{Q}_{12}^{(\kappa)}\, f_{12}\, |iI\rangle, \tag{6}$$

where $\left|\Psi^{(1)}_{\mathrm{OBS}}(t)\right\rangle$ contains the conventional OBS doubles, whose amplitudes are determined in the presence of the fixed geminals. The pair functions in the second term denote double excitations of the HF reference summed over both electronic spins. The full wavefunction through first order is the HF reference plus Eq. 6.

The projector is

$$\hat{Q}_{12} = (\mathbf{1} - \hat{O}_1)(\mathbf{1} - \hat{O}_2) - \hat{V}_1\hat{V}_2, \tag{7}$$

where $\hat{O}_\nu$ is the occupied projector for particle $\nu$ and $\hat{V}_\nu$ is the virtual projector for particle $\nu$ of the OBS. Within the finite RIBS, decomposing its identity operator as $\mathbf{1} = \hat{P} + \hat{A}'$ where $\hat{P}$ is the OBS projector and $\hat{A}'$ is the CABS complement and using $\hat{O} \subset \hat{P}$, Eq. 7 is equivalent to

$$\mathbf{1} - \hat{Q}_{12} = \hat{P}_1\hat{P}_2 + \hat{O}_1\hat{A}'_2 + \hat{A}'_1\hat{O}_2, \tag{8}$$

which is the form used in evaluating the F12 intermediates.

The amplitudes $t_{iI}^{(\kappa)}$ in Eq. 6 are evaluated in the SP style of Ten-no,[6,67]

$$t_{iI}^{(\kappa)} = t_{ep} \quad \text{for all } (i, I), \tag{9}$$

with the universal value $t_{ep}$ fixed by the electron-proton cusp condition. Because electrons and quantum protons are distinguishable, there is no analogue of the electron-electron singlet/triplet pair-symmetry split that motivated the two distinct fixed amplitudes in Ten-no's original SP ansatz. The electron-proton pair carries a single cusp value, and we correspondingly use a single universal $t_{ep}$ with diagonal structure in place of the singlet–triplet branched form. Analogously to the motivation for the introduction of the SP ansatz in single-component MP2-F12 theory, Eq. 9 makes the multicomponent MP2-F12 wavefunction orbital-invariant and removes any need for variational optimization of the amplitudes.

**II.D Electron-Proton Kato Cusp**

The generalized cusp condition of Pack and Brown for two non-relativistic particles of masses $m_1, m_2$ and charges $q_1, q_2$ interacting via the Coulomb potential is

$$\left\langle \frac{\partial \Psi}{\partial r} \right\rangle_{\text{sph}} \Big|_{r \to 0^+} = \mu\, q_1 q_2\, \Psi(r = 0), \tag{10}$$

where

$$\mu = \frac{m_1 m_2}{m_1 + m_2} \tag{11}$$

is the reduced mass and "sph" denotes spherical averaging.[4]

For two electrons ($\mu_{ee} = 1/2$, $q_e q_e = +1$), these equations give the canonical value, $t_{ee} = 1/2$, used in single-component F12. For an electron–proton pair ($m_e = 1$, $m_p = 1836.152673$, $q_e q_p = -1$),

$$\mu_{ep} = \frac{m_p}{1 + m_p} \approx 0.99946, \tag{12}$$

and so $|\text{cusp}| = \mu_{ep}$. In the SP fixed-amplitude ansatz, we take this magnitude as the geminal coefficient,

$$t_{ep} = \mu_{ep} = \frac{m_p}{1 + m_p} \approx 0.99946. \tag{13}$$

The signs of the electron-electron and electron-proton cusps are opposite. In the electron-electron case, the wave function has a positive cusp slope or Coulomb hole, while in the electron-proton case, the wave function has a negative cusp slope or Coulomb peak. This difference reflects the change of sign of $q_1 q_2$ in Eq. 10. The slope magnitude is $\mu$ in both cases and it is the magnitude that enters Eq. 6 as the geminal coefficient. The overall sign of $t_{ep}$ in our equations follows from the form of the geminal. With $f_{12}(r) = (1/\zeta)e^{-\zeta r}$ (Section III), whose slope at r = 0 is −1, the product $t_{ep} f_{12}$ carries the negative slope of the electron-proton cusp. We use this condition to fix the first-order geminal amplitude, with the slope referenced to the zeroth-order pair function. The finite Gaussian expansion used for the integrals approximates the Slater geminal but has zero radial derivative at the origin. Its adequacy is therefore assessed through convergence of the F12 energy with the fit length.

Because $m_p \gg m_e$, $\mu_{ep}$ is nearly twice $\mu_{ee}$. The electron-proton cusp is correspondingly about 2 times as steep as the electron-electron cusp. $t_{ep}$ is set by the physics of the cusp. Therefore, only the geminal length parameter $\zeta$ remains as a free parameter. In this work all quantum nuclei are protons ($m_\kappa = m_p$ for all $\kappa$), so the value of $t_{ep}$ in Eq. 6 is the same for all $\kappa$. If calculations were to be performed with heavier isotopes such as deuterium, these calculations would need to use the per-nucleus amplitude $t_{en}^{(\kappa)} = m_\kappa/(1 + m_\kappa)$, where *n* labels the isotopic particle. The corresponding mass must also be used in the reference kinetic operator, Eq. 22, and the CABS Fock matrices, along with the reference orbitals and energies corresponding to the particular nucleus and isotope.

**II.E Multicomponent MP2-F12 Intermediates**

We use the Standard Approximation C approach to evaluate the F12 intermediates.[68] We choose Standard Approximation C because its kinetic contribution is an RI-free double commutator that can be evaluated with the appropriate particle masses. The remaining Hartree, exchange, and projector terms are evaluated using finite-RIBS resolutions. With $g_{ep} = -r_{12}^{-1}$ as the electron-proton Coulomb operator and $\hat{Q}_{12}$ the projector from Eq. 8, the three working intermediates per pair $(i, I)$ are

$$V_{iI} \;=\; \langle iI|\, f_{12}\, \hat{Q}_{12}\, g_{ep}\, |iI\rangle, \tag{14}$$

$$X_{iI} \;=\; \langle iI|\, f_{12}\, \hat{Q}_{12}\, f_{12}\, |iI\rangle, \tag{15}$$

$$B_{iI} \;=\; B_{iI}^{\text{kin}} \;+\; B_{iI}^{\text{Hartree}} \;-\; B_{iI}^{\text{compl}}. \tag{16}$$

where the components of $B_{iI}$ are discussed below, with explicit working equations detailed in the Supporting Information. As a convention, the electron-proton Coulomb matrix elements carry the attractive sign that arises between negatively charged electrons and positively charged protons.

In practice, $V_{iI}$ is evaluated by expanding $\hat{Q}_{12}$ using Eq. 8 and requires evaluating a matrix element $\langle iI|f_{12}\, g_{ep}|iI\rangle$, a single explicitly correlated integral over the combined Slater-Coulomb operator, minus three integrals in which a single geminal R-integral,

$$R_{\alpha\beta}^{iI} \;\equiv\; \langle iI|f_{12}|\alpha\beta\rangle, \tag{17}$$

contracts against the conventional ep Coulomb two-particle integral $\langle\alpha\beta|g_{ep}|iI\rangle$, with $\alpha, \beta$ ranging over the three RIBS-block index pairs of Eq. 8: $(P_1, P_2)$ for the OBS–OBS subtraction, $(O_1, A'_2)$ for the occupied-electron/CABS-proton subtraction, and $(A'_1, O_2)$ for the CABS-electron/occupied-proton subtraction.

$X_{iI}$ has analogous structure to $V_{iI}$ with a matrix element $\langle iI|f_{12}^2|iI\rangle$ minus the three RI terms in which two geminal integrals contract.

The decomposition of $B_{iI}$ in Eq. 16 is more complicated than $V_{iI}$ or $X_{iI}$. First, we introduce notation. Let $\hat{F}^{\nu}$ and $\hat{K}^{\nu}$ denote the one-particle Fock and exchange operators for particle $\nu \in \{e, p^{\kappa}\}$ over the RIBS basis, $\hat{F}^{\nu}_{PQ} = \langle P^{\nu}|\hat{f}^{\nu}|Q^{\nu}\rangle$ and $\hat{K}^{\nu}_{PQ} = \langle P^{\nu}|\hat{k}^{\nu}|Q^{\nu}\rangle$ with *P*, *Q* running over occupied, OBS-virtual, and CABS blocks. We note that $K^{p^{\kappa}} = 0$ because protons are treated as distinguishable and carry no exchange. The protonic Fock operator $\hat{f}^{p^{\kappa}}$ is the proton kinetic energy plus the mean Coulomb field of the electrons, the mean Coulomb field of the classical nuclei, and the mean Coulomb field of every other quantum proton. This matches the multicomponent HF reference protonic Fock operator for which the conventional multicomponent MP2 energy is defined. For a single quantum nucleus, the last term is absent. Finally, we also define the Hartree (*i.e.*, exchange-free) Fock operator $\hat{F}_{\mathrm{H}} = \hat{F} + \hat{K}$.

The kinetic double-commutator piece $B_{iI}^{\mathrm{kin}}$ is evaluated explicitly and is discussed more below. The Hartree contractions $B_{iI}^{\mathrm{Hartree}}$ arise from the $f_{12}^2\hat{F}_{\mathrm{H}}$ trace, and the completeness corrections $B_{iI}^{\mathrm{compl}}$ contain diagonal and off-diagonal Fock terms and the electronic exchange correction. Additional Fock couplings would be removed by the generalized and extended Brillouin conditions[65] (GBC and EBC, respectively). In this study, we do not impose either the GBC or EBC. Therefore, all eleven RIBS-block contractions over the occupied, OBS-virtual, and CABS blocks are included, six for the electrons, namely five contractions of the form $R \cdot F \cdot R$ and one of the form $R \cdot K \cdot R$,[13,65,66] and five of the form $R \cdot F \cdot R$ for the proton, for which the exchange is zero. Each dot-product here denotes a tensor contraction over the inner RIBS slots of the two R-integrals, with *F* or *K* acting between two RIBS indices on a single particle while the bra- and ket- of each R-integral remain pinned to the diagonal occupied pair $(i, I)$.

The kinetic piece of **B** is the diagonal matrix element of the symmetrized double commutator of the two kinetic operators with the correlation factor,

$$B_{iI}^{\mathrm{kin}} = \frac{1}{2}\left\langle iI\right|\left[f_{12}, \left[\hat{T}_1 + \hat{T}_2, f_{12}\right]\right]\left|iI\right\rangle. \tag{18}$$

For a one-parameter geminal $f_{12} = f(r)$ with $r = |\mathbf{r}_1 - \mathbf{r}_2|$ and one-particle kinetic operators $\hat{T}_\nu = -(1/2m_\nu)\nabla_\nu^2$,

$$\left[\hat{T}_\nu, f\right] = -\frac{1}{2m_\nu}(\nabla_\nu^2 f) - \frac{1}{m_\nu}(\nabla_\nu f)\cdot\nabla_\nu. \tag{19}$$

The outer commutator with f removes the multiplicative Laplacian term and retains the contribution from the gradient operator, which gives

$$\left[f, \left[\hat{T}_\nu, f\right]\right] = \frac{1}{m_\nu}(\nabla_\nu f)^2 = \frac{1}{m_\nu}|\nabla f|^2, \tag{20}$$

where $\nabla_1 f = +(\partial_r f)\,\hat{r}$ and $\nabla_2 f = -(\partial_r f)\,\hat{r}$ give the same magnitude for both particles. Summing over the two particles,

$$\left[f, \left[\hat{T}_1 + \hat{T}_2, f\right]\right] = \left(\frac{1}{m_1} + \frac{1}{m_2}\right)|\nabla f|^2, \tag{21}$$

which gives the final result,

$$B_{iI}^{\mathrm{kin}} = \frac{1}{2}\left(\frac{1}{m_e} + \frac{1}{m_p}\right)\langle iI||\nabla f|^2\,|iI\rangle. \tag{22}$$

For two electrons ($m_1 = m_2 = 1$) the prefactor in Eq. 22 is $\frac{1}{2}(1+1) = 1$, recovering the canonical electron-electron result. For the electron-proton case,

$$\frac{1}{2}\left(1 + \frac{1}{m_p}\right) = \frac{1}{2}(1 + 5.446 \times 10^{-4}) \approx 0.500272. \tag{23}$$

The electron-proton kinetic prefactor is approximately one half of the electron-electron value. The larger cusp amplitude enters the energy separately through $t_{ep}{}^2$ in Eq. 30, whereas the squared gradient of $f(r) = \frac{e^{-\zeta r}}{\zeta}$ is $e^{-2\zeta r}$ . The relative magnitudes of the B intermediates therefore also depend on $\zeta$ and the occupied orbitals.

We evaluate the gradient integral in Eq. 22 directly using the integral library libint2,[69] which returns $\left\langle iI\left||\nabla\tilde{f}|^2\right|iI\right\rangle$ for the contracted Gaussian-type geminal fit $\tilde{f}(r) = \sum_k c_k\, e^{-\alpha_k r^2}$ of the Slater

exponential $e^{-\zeta r}$. This integral is divided by $\zeta^2$ to convert from the bare-Slater convention to the $\left(\frac{1}{\zeta}\right) e^{-\zeta r}$ geminal form that we assume for the geminal function.

The **V** intermediate describes the linear reference-geminal contribution, while **B** and **X** describe the quadratic geminal contribution. They are not, however, the complete second-order F12 correction because the full first-order wavefunction contains the conventional multicomponent MP2 doubles in addition to the geminal term of Eq. 6. Therefore, the second-order F12 energy includes cross-terms in which the F12 geminal contracts with the conventional MP2 amplitudes. These cross-terms vanish only if the off-diagonal virtual-CABS Fock matrix elements $F_a^{a'}$ (electron) and $F_A^{A'}$ (proton) vanish. As they do not in general, we retain them in our expressions.

The multicomponent MP2-F12 cross-term contributions are then encoded in the *c*-intermediate,

$$c_{aA}^{iI} = \sum_{a' \in \mathrm{CABS}_e} R_{a'A}^{iI} F_a^{a'} + \sum_{A' \in \mathrm{CABS}_p} R_{aA'}^{iI} F_A^{A'}, \tag{24}$$

where $a'$ runs over the electronic CABS, $A'$ runs over the protonic CABS, and the two **F** matrices are the corresponding electronic and protonic Fock matrices. The *c*-intermediate itself does not enter the conventional multicomponent MP2 energy of Eq. 3. Instead, it appears in the total multicomponent MP2-F12 energy only through the V- and B-coupling intermediates,

$$v_{iI}^{\mathrm{coup}} = \sum_{aA} \frac{c_{aA}^{iI} \langle iI \mid g_{ep} \mid aA \rangle}{\epsilon_i^e + \epsilon_I^p - \epsilon_a^e - \epsilon_A^p}, \tag{25}$$

$$b_{iI}^{\mathrm{coup}} = \sum_{aA} \frac{\left(c_{aA}^{iI}\right)^2}{\epsilon_i^e + \epsilon_I^p - \epsilon_a^e - \epsilon_A^p}. \tag{26}$$

Therefore, the *c*-intermediate couples the F12 amplitudes to the conventional MP2 doubles amplitudes through Eqs. 25 and 26, which enter the electron-proton F12 energy correction, $\Delta E_{F12}^{ep^{\kappa}}$, in Eq. 30 below with prefactors $2n^e t_{ep}$ and $n^e t_{ep}^2$, respectively. Without these terms, the resulting electron-proton F12 correction would not be a proper second-order energy.[14,66]

### II.F CABS Singles Correction

The Hartree-Fock basis-set error is corrected separately for the electrons and for each of the quantum protons via the CABS singles expression.[15,70] For the electrons, the singles amplitudes $t_i^{a'}$ solve the equation,

$$-\hat{F}^{\mathrm{occ}}\, t \;+\; t\,\hat{F}^{\mathrm{CABS}} \;=\; -\hat{F}^{\mathrm{occ,CABS}}, \tag{27}$$

and the energy contribution, $\Delta E_S^e$, for the electrons is

$$\Delta E_S^e \;=\; n^e \sum_{ia'} t_i^{a'}\, F_i^{a'}, \tag{28}$$

with the spin-counting prefactor $n^e = 2$ for the closed-shell electrons. The amplitudes $t_i^{a'}$ span the OBS virtual and CABS orbitals, so that a′ in Eqs. 27 and 28 runs over both sets. The equations for distinguishable quantum protons are analogous to the electronic case except the sum is over *I* and $A'_p$ with $n^p = 1$.

### II.G Multicomponent MP2-F12 Total Energy

Collecting the contributions, the multicomponent MP2-F12 total energy is

$$E_{\text{MC-MP2-F12}} \;=\; E_{\text{MC-HF}} \;+\; E_{\mathrm{MP2}}^{ee} \;+\; \sum_{\kappa=1}^{N_q}\left[E_{\mathrm{MP2}}^{ep^\kappa} \;+\; \Delta E_{F12}^{ep^\kappa} \;+\; \Delta E_S^{p^\kappa}\right] \;+\; \Delta E_S^e, \tag{29}$$

with

$$\Delta E_{F12}^{ep^\kappa} \;=\; n^e\left\{\sum_{iI}\left[2\,t_{ep}\left(V_{iI} + v_{iI}^{\mathrm{coup}}\right) + t_{ep}^2\left(B_{iI} + b_{iI}^{\mathrm{coup}}\right)\right] \;-\; t_{ep}^2\, X_F^{(\kappa)}\right\}, \tag{30}$$

where $n^e = 2$ for closed-shell systems, the sum runs over *I* on nucleus $\kappa$, and

$$X_F^{(\kappa)} \;=\; \sum_{iI}\left[\sum_k X_{iI,kI}\, F_{ki}^e \;+\; \sum_K X_{iI,iK}\, F_{IK}^{p^\kappa}\right], \tag{31}$$

where $X_{iI,kK} = \langle iI| f_{12}\, \hat{Q}_{12}\, f_{12}\, |kK\rangle$ is the off-diagonal generalization of Eq. (15). For canonical multicomponent HF orbitals, the occupied Fock blocks are diagonal, $F^{e}_{ki} = \epsilon^{e}_{i}\, \delta_{ki}$ and $F^{p^\kappa}_{IK} = \epsilon^{p^\kappa}_{I}\, \delta_{IK}$, and Eq. (31) reduces to the Hylleraas penalty $X^{(\kappa)}_{F} = \sum_{iI} \left(\epsilon^{e}_{i} + \epsilon^{p^\kappa}_{I}\right) X_{iI}$ similar to single-component MP2-F12 theory.

In the following, we denote the total CABS singles correction summed over all protonic and electronic species as $\Delta E_S$ .

For the reader familiar with F12 methodology, we highlight the similarity of the working equations in Section II to the working equations for correcting the electron-electron correlation energy in single-component MP2-F12 theory. This similarity is one of the most appealing features of multicomponent methods and facilitates their derivation and implementation.

**III. Implementation and Computational Details**

All calculations were performed with a multicomponent MP2-F12 code built on top of the PySCF framework.[71] The integral library libint2 was used to calculate all explicitly correlated integrals.[69]

The electron-proton F12 correction has no external reference implementation, so we validate it in two steps. First, when the second particle is assigned the electron mass and charge and the electronic basis sets, the electron-proton **V**, **X**, and coupling intermediates reproduce the electron-electron intermediates of Standard Approximation C to better than $5 \times 10^{-14}$ Ha, and **B** does so to $3 \times 10^{-12}$ Ha once the proton exchange term, which is absent for a distinguishable particle, is removed (Be atom, cc-pVDZ basis set). Second, the electron-electron MP2-F12 correlation energy computed with the same working equations reproduces the MPQC program[72] to $5.6 \times 10^{-9}$ Ha for $H_2O$ with the cc-pVDZ basis set (Standard Approximation C, with the kinetic double commutator evaluated with the explicit gradient integral of Eq. 22). These comparisons test the electronic limit of the working equations. As an internal check of the multi-proton bookkeeping, the two symmetry-equivalent protons of HCCH and HNNH give per-proton corrections that agree to about $10^{-14}$ Ha.

The electron-spin factor, $n^e$, of the electron-proton F12 correction was validated separately. Because the quantum proton forms a correlation cusp with both the alpha and beta electrons of each doubly occupied spatial orbital, every spatial pair (i,I) contributes two geminal spin channels to the wavefunction of Eq. 6, each with the cusp amplitude t_ep, giving the overall prefactor $n^e = 2$. We confirmed this factor with an exactly solvable second-quantized model consisting of three electronic and three protonic orbitals, for which the Hylleraas functional of the ansatz of Eq. 6 was evaluated by explicit matrix algebra in the full Fock space, with the conventional doubles amplitudes minimized exactly.

The model reproduces the conventional electron-proton MP2 energy of Eq. 5, including its electron-spin factor of two, as well as the validated electron-electron diagonal-pair energy convention, and confirms the prefactor $n^e = 2$ in Eq. 30 uniformly in the **V**, **B**, **X**, and coupling terms. The factor was additionally confirmed by an independent symbolic derivation with the SASQ code,[73] in which the proton is represented by a fermionic operator that commutes with all electronic operators. The numerical model and the derived equations (Eqs. S1-S5) are given in Section S.III of the Supporting Information.

For the OBS, multicomponent MP2-F12 calculations were performed using the augmented correlation-consistent series of basis sets aug-cc-pVXZ (X=D,T,Q).[74] The aug-cc-pV5Z basis set was used for the near complete basis set (CBS) conventional (i.e., no-F12) multicomponent MP2 reference, except for $H_5O_2^+$ where aug-cc-pV5Z was computationally infeasible and an aug-cc-pV5Z density-fitted conventional multicomponent MP2 calculation was used as the CBS proxy. The protonic OBS is PB4-D, which was shown to be sufficient when compared to the PB4-F1 protonic basis set as discussed below.[75] All electrons were correlated in the multicomponent MP2 and MP2-F12 calculations.

As previously mentioned, we use Ansatz 3 of Werner, Adler, and Manby with the SP fixed amplitude $t_{ep} = \mu_{ep} \approx 0.99946$ and Standard Approximation C for the **B** and **X** intermediates. The Slater geminal, $f_{12}(r) = (1/\zeta)e^{-\zeta r}$, uses $\zeta = 1.7$ bohr$^{-1}$, which was determined to be the best global value across the test set with the aug-cc-pV5Z-RI ABS used for the calculations in Table 1, as shown in Section IV.B. The Slater geminal is fit by 20 contracted Gaussian-type geminals with exponents in a geometric progression on $[0.02\zeta^2, 5000\zeta^2]$ and coefficients from $r^2$-weighted least-squares minimization on $r \in [10^{-5}, 15/\zeta]$ on a 4000-point grid. The 20-term fit was verified to converge the F12 electron-proton correction to within 2×$10^{-6}$ Ha of a 24-term reference (Section IV.B). The squared geminal is fit separately to $\frac{e^{-2\zeta r}}{\zeta^2}$. For HCN with aug-cc-pVDZ and the auxiliary bases used in Table 1, replacing this fit by the exact product of the 20-Gaussian geminal fit changes the F12 correction by −1.20 μHa (0.017%).

In our MP2-F12 implementation, the GBC and EBC are not assumed. Therefore, taking all of our approximations together and using the standard F12 nomenclature, the present method is the electron–proton analogue of MP2-F12/3C(FIX) as we use Ansatz 3, approximation C for the **B**/**X** intermediates and a fixed cusp-determined geminal amplitude.[6,66] Because neither the EBC nor the GBC is invoked, it is the unstarred "3C" rather than the EBC-based 3*C variant. We use multicomponent MP2-F12 as the name of the new method throughout.

We use the CABS+ formulation with the RIBS taken as the union of the OBS and a fixed ABS. For all calculations, the electronic ABS is aug-cc-pV5Z-RI,[76] which was chosen as it is a sufficient ABS for every OBS in this study. The ABS is thus held fixed across the electronic OBS series, with auxiliary-basis convergence examined in Section IV and the Supporting Information. Using a larger cardinality ABS than the corresponding OBS has also been previously shown to lower errors in density-fitted multicomponent CCSD calculations.[40] As discussed in Section IV, we also investigated using the OptRI family of basis sets[77] for the ABS but found them to be less accurate when computing the $\Delta E_S$ CABS correction.

We note that aug-cc-pV5Z-RI is formally the RI-MP2 density-fitting auxiliary, which is one of the three distinct auxiliary roles in modern F12 theory: Coulomb and/or exchange fitting (JKFIT), correlation density fitting (RI-MP2/MP2Fit), and the F12 one-particle complement (OptRI/CABS).[78] In this study the RI-MP2 basis set also acts as the CABS. A saturated RI-MP2 auxiliary reproduces the dedicated OptRI CABS to within a few hundredths of a kcal/mol at the TZ and QZ levels, with a residual RI error an order of magnitude smaller than the basis-set incompleteness error itself, as shown by Semidalas and Martin.[78] The aug-cc-pVnZ-OptRI ABSs are available,[77] but the OptRI basis sets do not perform well for the $\Delta E_S$ CABS correction, which is known in the literature, as this deficiency motivated the introduction of the augmented OptRI+ ABSs.[79] We adopt the singles-converging RI-MP2 auxiliary and benchmark it directly against the CABS-independent Hartree-Fock basis-set error below.

The protonic ABS is an even-tempered set of three primitives per angular momentum, 3s3p3d, with the exponents chosen as an even-tempered progression of $(0.25, 4.47, 80.0)$ bohr$^{-2}$. These protonic ABS basis functions are centered on each quantum hydrogen. Reducing the protonic ABS to this size changes the electron-proton F12 correction by less than 0.05 %, as discussed below.

The CABS coefficients are constructed from each ABS by an eigenvalue-threshold canonical orthonormalization of the RIBS overlap. Eigenvalues below a relative threshold $\lambda < 10^{-8} \cdot \lambda_{\max}$, where $\lambda_{\max}$ is the largest eigenvalue, are discarded. The CABS is then obtained by diagonalization of the OBS-orthogonal projector $\hat{Q} = \mathbf{1} - \hat{M}^{\top}\hat{M}$ where $\hat{M} = \mathbf{C}_{\mathrm{OBS}}^{T}\mathbf{S}\widetilde{\mathbf{X}}$ is the overlap between the OBS molecular orbitals and the canonically orthonormalized RIBS functions. The CABS space is then chosen as the set of eigenvectors with eigenvalue > 0.5. We use this procedure instead of the index-split singular-value decomposition of the original CABS construction. The relative rather than absolute linear-dependence cutoff is essential for diffuse augmented ABSs, for which the overlap has large collective eigenvalues, because an absolute cutoff would retain near linearly-dependent auxiliary functions that the relative cutoff removes. Finally, we note that the electron-proton F12 matrix elements are resolved over the joint RIBS in which each particle carries either an electronic or protonic ABS. In practice, the two per-particle RIBSs are concatenated into a single combined molecular basis, and the required cross electron-proton integrals are evaluated as the appropriate OBS/CABS shell-blocks of that combined basis.

The $\Delta E_S$ CABS correction is included separately for the electrons and for each quantum nucleus, with the corresponding amplitudes obtained from Eq. 27.

To reduce the memory overhead, the construction of the electronic Coulomb (**J**) and exchange (**K**) matrices over the CABS space was performed with density fitting using the Coulomb metric and the aug-cc-pVQZ-jkfit auxiliary basis.[80] The residual density-fitting error is below $10^{-6}$ Ha on $\Delta E_{F12}^{ep}$ and below $10^{-5}$ Ha on $\Delta E_S$ (Section IV.B). The electron-proton F12 geminal integrals and the libint2 $\left\langle iI \left| \left|\nabla \tilde{f}\right|^2 \right| iI \right\rangle$ integrals were calculated without density fitting as tests indicated that density-fitting these integrals through the CABS amplifies the density fitting error. The multicomponent MP2 reference values at aug-cc-pV5Z are computed without any density fitting, except for $H_5O_2^+$ for which density-fitted multicomponent MP2 was used (Section IV.A).

For all calculations, molecular geometries were obtained from a single-component MP2 geometry optimization using the aug-cc-pVTZ electronic basis set with no density fitting. These geometries were used for the F12 calculations for all other basis sets to allow the convergence of the electron-proton correlation energy with respect to the electronic basis set to be tested. For $H_5O_2^+$, we constrained it to the $C_2$-symmetric Zundel structure during the geometry optimization. All geometries used in this work are tabulated in the Supporting Information.

## IV. Results and Discussion

### IV.A Benchmarking

We benchmark the multicomponent MP2-F12 method on HCN, $FHF^-$, HCCH, HNNH, and $H_5O_2^+$. For the first four molecules, we treat all hydrogen nuclei quantum mechanically. For $H_5O_2^+$, we treat the bridging hydrogen nucleus quantum mechanically. These five molecules are chosen to span neutral, anionic, and cationic systems, terminal and bridging quantum protons, and one versus two quantum nuclei. We focus on the electron-proton correlation energy and its convergence toward the CBS limit. We compute the conventional non-F12 electron-proton MP2 energy, $E_{\mathrm{MP2}}^{ep} = \sum_{\kappa} E_{MP2}^{ep^{\kappa}}$, the F12 correction, $\Delta E_{F12}^{ep} = \sum_{\kappa} \Delta\, E_{F12}^{ep^{\kappa}}$, and their sum, $E_{\mathrm{MP2}}^{ep}$+ $\Delta E_{F12}^{ep}$, which is the electron-proton MP2-F12 correlation energy. We report these alongside the one-particle $\Delta E_S$ CABS correction.

The finite-basis reference for the total electron-proton MP2 correlation energy is the $E_{\mathrm{MP2}}^{ep}$ value calculated with the aug-cc-pV5Z basis set. For $H_5O_2^+$, where a canonical multicomponent MP2 calculation with aug-cc-pV5Z is computationally infeasible, the aug-cc-pV5Z $E_{\mathrm{MP2}}^{ep}$ value was obtained with density-fitted multicomponent MP2. A test calculation on $H_5O_2^+$ with the aug-cc-pVQZ electronic basis set using the canonical and density-fitted multicomponent MP2 codes differed by less than 1 μHa. Because the reference is the aug-cc-pV5Z $E_{\mathrm{MP2}}^{ep}$ value rather than an extrapolated CBS value, total MP2-F12 correlation energies at or slightly above 100% indicate that the F12 result has reached the aug-cc-pV5Z level although this comparison alone does not determine the error relative to the true CBS limit.

Table 1 shows the electron-proton correlation energies for the five molecules. These calculations use the aug-cc-pV5Z-RI ABS for every OBS as discussed more in Section IV.B. The convergence of the MP2-F12 electron-proton correlation energy with respect to the auxiliary-basis convention is discussed more in the Supporting Information. $E_{\mathrm{MP2}}^{ep}$ converges slowly, recovering 59–61% of the aug-cc-pV5Z value at aug-cc-pVDZ, 79–82% at aug-cc-pVTZ, and 91–93% at aug-cc-pVQZ. The increments between successive cardinals decrease more slowly than the $X^{-3}$ behavior of the electron-electron correlation energy, so the electron-proton correlation energy cannot be extrapolated with the same confidence as in the single-component case. A two-point $X^{-3}$ extrapolation from the aug-cc-pVQZ and aug-cc-pV5Z values places the aug-cc-pV5Z energy at 91–93% of the estimated basis-set limit, whereas the same extrapolation from the aug-cc-pVTZ and aug-cc-pVQZ values places it at 99–100%. We use the (Q,5) estimate for the comparisons below. Changing the extrapolation pair to (T,Q) shifts the estimated limits by 6.4–8.8%, or 0.83–2.97 mHa, across the five systems. Both extrapolations concern the electronic basis with PB4-D held fixed.

Adding the F12 correction brings every cardinality of the electronic basis set close to the (Q,5) estimate. Measured against the (Q,5) estimate, multicomponent MP2-F12 recovers 96.0–97.7% at aug-cc-pVDZ, 97.4–99.2% at aug-cc-pVTZ, and 98.9–100.3% at aug-cc-pVQZ, a nearly cardinality-independent recovery of the kind familiar from single-component MP2-F12. Relative to the finite aug-cc-pV5Z reference itself, the corrected energies lie at 103.5–106.6%, 105.0–108.3%, and 106.6–109.7% (Table 1).

This near-uniform recovery of the estimated CBS limit at every cardinality is consistent across all five systems, although the inferred fraction of missing correlation recovered by the single-cusp SP ansatz depends on the extrapolated reference. The residual few-percent deviation from the CBS limit at aug-cc-pVDZ may reflect the incompleteness of the fixed-zeta single-geminal ansatz and of the DZ orbital basis for the non-cusp part of the correlation, but the present calculations do not separate these effects from extrapolation error.

| System | OBS | $E_{\mathrm{MP2}}^{ep}$ | $\Delta E_{F12}^{ep}$ | $E_{\mathrm{MP2}}^{ep} + \Delta E_{F12}^{ep}$ | $E_{\mathrm{MP2}}^{ep}$ (%) | $E_{\mathrm{MP2}}^{ep} + \Delta E_{F12}^{ep}$ (%) |
|---|---|---|---|---|---|---|
| HCN | aDZ | −0.009083 | −0.006947 | −0.016030 | 60.3 | 106.4 |
| | aTZ | −0.011969 | −0.004304 | −0.016273 | 79.5 | 108.0 |
| | aQZ | −0.013782 | −0.002672 | −0.016454 | 91.5 | 109.2 |
| | a5Z | −0.015064 | | | 100 | |
| $FHF^-$ | aDZ | −0.007273 | −0.005261 | −0.012534 | 60.1 | 103.5 |
| | aTZ | −0.009796 | −0.002920 | −0.012716 | 80.9 | 105.0 |
| | aQZ | −0.011201 | −0.001706 | −0.012907 | 92.5 | 106.6 |
| | a5Z | −0.012106 | | | 100 | |
| $H_5O_2^+$ | aDZ | −0.007470 | −0.005239 | −0.012709 | 60.9 | 103.6 |
| | aTZ | −0.009994 | −0.002947 | −0.012942 | 81.5 | 105.5 |
| | aQZ | −0.011369 | −0.001731 | −0.013100 | 92.7 | 106.8 |
| | a5Z | −0.012269 | | | 100 | |
| HCCH | aDZ | −0.018504 | −0.014143 | −0.032648 | 60.3 | 106.4 |
| | aTZ | −0.024374 | −0.008769 | −0.033142 | 79.4 | 108.0 |
| | aQZ | −0.028072 | −0.005445 | −0.033517 | 91.5 | 109.2 |
| | a5Z | −0.030684 | | | 100 | |
| HNNH | aDZ | −0.018214 | −0.014791 | −0.033005 | 58.8 | 106.6 |
| | aTZ | −0.024360 | −0.009195 | −0.033555 | 78.7 | 108.3 |
| | aQZ | −0.028167 | −0.005798 | −0.033965 | 90.9 | 109.7 |
| | a5Z | −0.030971 | | | 100 | |

**Table 1:** Electron-proton correlation energy (Ha) and fraction of the finite-basis ($E_{\mathrm{MP2}}^{ep}$, aug-cc-pV5Z) reference recovered, without and with the F12 correction. For the two-proton systems the energies are summed over both protons. Calculations are performed with the aug-cc-pVXZ (aXZ) electronic basis sets. The $E_{\mathrm{MP2}}^{ep} + \Delta E_{F12}^{ep}$ column may differ from the sum of its two preceding columns in the last digit due to rounding.

As shown in Table 2, the $\Delta E_S$ CABS correction is dominated by the electronic contribution. The protonic $\Delta E_S$ CABS corrections are of order $10^{-5}$ Ha or smaller per proton and are negligible by comparison to the electronic $\Delta E_S$ CABS correction, which is consistent with the PB4-D protonic OBS being well converged. The $\Delta E_S$ CABS correction is large at aug-cc-pVDZ and is largest for the diffuse anionic $FHF^-$ and cationic $H_5O_2^+$ systems. However, the $\Delta E_S$ CABS corrections reduce rapidly, decreasing by roughly a factor of three to five per cardinal, as the OBS saturates the occupied-orbital HF limit. At aug-cc-pVDZ, the $\Delta E_S$ CABS correction exceeds the $\Delta E_{F12}^{ep}$ in magnitude, so both pieces are required for a balanced small-basis result, but by aug-cc-pVQZ, the $\Delta E_S$ CABS corrections have fallen to a magnitude on the order of $10^{-3}$ Ha.

| | | Electronic $\Delta E_S$ | Protonic $\Delta E_S$ | Total $\Delta E_S$ |
|---|---|---|---|---|
| HCN | aDZ | −0.023247 | −0.000013 | −0.023260 |
| | aTZ | −0.005049 | −0.000013 | −0.005062 |
| | aQZ | −0.001254 | −0.000013 | −0.001267 |
| $FHF^-$ | aDZ | −0.060868 | −0.000001 | −0.060869 |
| | aTZ | −0.013776 | −0.000001 | −0.013778 |
| | aQZ | −0.003096 | −0.000001 | −0.003098 |
| $H_5O_2^+$ | aDZ | −0.048255 | −0.000002 | −0.048257 |
| | aTZ | −0.010156 | −0.000002 | −0.010157 |
| | aQZ | −0.002165 | −0.000001 | −0.002166 |
| HCCH | aDZ | −0.021367 | −0.000028 | −0.021395 |
| | aTZ | −0.004648 | −0.000027 | −0.004675 |
| | aQZ | −0.001532 | −0.000027 | −0.001559 |
| HNNH | aDZ | −0.025565 | −0.000033 | −0.025598 |
| | aTZ | −0.006647 | −0.000030 | −0.006677 |
| | aQZ | −0.001911 | −0.000032 | −0.001943 |

**Table 2:** $\Delta E_S$ CABS correction (Ha). Calculations are performed with the aug-cc-pVXZ (aXZ) electronic basis sets.

Because the $\Delta E_S$ CABS corrections are a second-order estimate of the one-particle HF basis-set error, we benchmark them directly against that error, defined as the difference between the multicomponent HF energies using the OBS versus using the aug-cc-pV5Z reference basis set, using density fitting for $H_5O_2^+$ as specified in Table 3, as shown in Table 3. The $\Delta E_S$ CABS corrections recover 75–90% of the HF basis-set error for the aug-cc-pVDZ and aug-cc-pVTZ basis sets across the test set. For the aug-cc-pVQZ basis set, the percentages are not meaningful because the remaining HF basis-set error is only 1–4 mHa and the aug-cc-pV5Z reference itself is not converged at this level. Achieving this level of agreement requires an ABS that spans the one-particle complement into which the singles relax, and the aug-cc-pV5Z-RI auxiliary does so.

In contrast, the OptRI family of basis sets recovers only 23–49% of the HF basis-set error (Table 6) and is unsuitable for the $\Delta E_S$ CABS correction in multicomponent MP2, as discussed below. This deficiency is the central reason the ABS is aug-cc-pV5Z-RI rather than an OptRI set.

| | OBS | Total $\Delta E_S$ | HF basis error | % Recovered |
|---|---|---|---|---|
| HCN | aDZ | −0.023260 | −0.026265 | 89% |
| | aTZ | −0.005062 | −0.006691 | 76% |
| | aQZ | −0.001267 | −0.001216 | 104% |
| $FHF^-$ | aDZ | −0.060869 | −0.067523 | 90% |
| | aTZ | −0.013778 | −0.017572 | 78% |
| | aQZ | −0.003098 | −0.003770 | 82% |
| $H_5O_2^+$ | aDZ | −0.048257 | −0.054644 | 88% |
| | aTZ | −0.010157 | −0.013563 | 75% |
| | aQZ | −0.002166 | −0.002705 | 80% |
| HCCH | aDZ | −0.021395 | −0.023810 | 90% |
| | aTZ | −0.004675 | −0.005636 | 83% |
| | aQZ | −0.001559 | −0.001121 | 139% |
| HNNH | aDZ | −0.025598 | −0.028746 | 89% |
| | aTZ | −0.006677 | −0.008341 | 80% |
| | aQZ | −0.001943 | −0.001616 | 120% |

**Table 3:** $\Delta E_S$ CABS correction (Ha) benchmarked against the multicomponent HF energy correction $E_{\mathrm{HF}}[5\mathrm{Z}] - E_{\mathrm{HF}}[\mathrm{OBS}]$. Calculations are performed with the aug-cc-pVXZ (aXZ) electronic basis sets using the aug-cc-pV5Z electronic basis set as the finite-basis reference. The HF energies are multicomponent HF energies with the PB4-D protonic basis set; for $H_5O_2^+$ the aug-cc-pV5Z value is density fitted with the aug-cc-pV5Z-jkfit auxiliary basis set.

For the two-proton systems, the per-proton F12 contributions, $\Delta E_{F12}^{ep^\kappa}$, are equal by nearly machine precision as HCCH and HNNH each have two symmetry-equivalent terminal protons, and $\Delta E_{F12}^{ep^{\kappa_1}} = \Delta E_{F12}^{ep^{\kappa_2}}$ with an asymmetry below $10^{-12}$ Ha for every OBS. The convergence behavior is the same as for the single-proton systems, which confirms that the per-nucleus, distinguishable-proton construction extends to several quantum nuclei.

## IV.B Calibration and Parameterization

The parameters given in Section III such as the geminal exponent $\zeta$ and its Gaussian-fit length and the electronic and protonic auxiliary bases were calibrated by single-parameter sweeps on HCN and $FHF^-$, with the geminal exponent also scanned for the full test set (Table 5). In this section, we summarize those calibrations.

The Slater geminal is represented by a fit to contracted Gaussian-type geminals. Table 4 shows $\Delta E_{F12}^{ep}$ versus the number of fit terms. For the geometric fits of Section III, in which the squared geminal is fit separately with the same number of terms, $\Delta E_{F12}^{ep}$ only stabilizes once the $r^2$-weighted fit residual falls below $\sim 10^{-4}$. We note that the fit residual is intrinsic to the geometric contracted Gaussian-type-geminal fit and is independent of system and basis. The 20-term fit used in this study converges $\Delta E_{F12}^{ep}$ to $\sim 2 \times 10^{-6}$ Ha of the 24-term reference. Therefore, we use the 20-term fit for the benchmarking calculations. Table 4 also includes the 6-term fit of Tew and Klopper,[81] which is widely used for the electron-electron geminal, refit at the geminal exponent used in this work. If the squared geminal is formed as the square of this 6-term expansion, the electron-proton F12 correction agrees with the 20-term result to within 2 μHa for HCN and 6 μHa for $FHF^-$. If the squared geminal is instead fit separately with 6 terms, the correction overshoots the 20-term result by 10% for HCN and 47% for $FHF^-$. Fitting the squared geminal separately with 20 terms, which reproduces it far more accurately than the square of the 6-term expansion does, still leaves the correction 1.3% below the 20-term result for HCN and 8.7% below for $FHF^-$. It is therefore the consistency of the squared geminal with the geminal, rather than the accuracy of its fit, that matters, and a 6-term fit is adequate for the electron-proton correction provided that the squared geminal is formed as its square.

| # of Terms | HCN | $FHF^-$ | Fit Residual |
|---|---|---|---|
| 6[a] | −0.007632 | −0.007711 | $3.5 \times 10^{-2}$ |
| 6[b] | −0.006948 | −0.005267 | $3.5 \times 10^{-2}$ |
| 10 | −0.007086 | −0.006546 | $1.9 \times 10^{-2}$ |
| 14 | −0.006936 | −0.005096 | $9.8 \times 10^{-4}$ |
| 16 | −0.006968 | −0.005311 | $2.8 \times 10^{-4}$ |
| 20 | −0.006946 | −0.005261 | $2.1 \times 10^{-5}$ |
| 24 | −0.006947 | −0.005263 | $1.4 \times 10^{-6}$ |

**Table 4**: $\Delta E_{F12}^{ep}$ (Ha) versus the number of Gaussians in the geminal fit with the aug-cc-pVDZ OBS and the aug-cc-pV5Z-RI ABS, together with the $r^2$-weighted (radial-volume) relative fit residual. The values in this table are from a separate calibration run and can differ from the values of Table 1 in the final digit (differences of less than 1 μHa). The 6-term rows use the fit of Tew and Klopper,[81] refit at the geminal exponent used in this work, with the squared geminal either fit separately with 6 terms (a) or formed as the square of the 6-term expansion (b). All other rows use the geometric fit of Section III. The fit residual is that of the geminal and is therefore the same for both 6-term rows; the corresponding residuals of the squared geminal are $6.2 \times 10^{-3}$ (a) and $1.5 \times 10^{-3}$ (b). For the geometric fits, the two residuals are identical.

$\Delta E_{F12}^{ep}$ is a broad, shallow function of $\zeta$ (Table 5). With the aug-cc-pV5Z-RI ABS, the optimal value of $\zeta$ is system dependent, lying at 1.3–1.6 bohr$^{-1}$ at the aug-cc-pVDZ level (Table 5) and at 1.6 (HCN) to 2.0 ($FHF^-$) bohr$^{-1}$ at the aug-cc-pVTZ level (Table S3). We choose $\zeta$ = 1.7 bohr$^{-1}$, the value that minimizes the largest deviation of $\Delta E_{F12}^{ep}$ from its $\zeta$-optimal value across both data sets. This deviation is 1.6%.

| $\zeta$ | HCN | $FHF^-$ | HCCH | $H_5O_2^+$ | HNNH |
|---|---|---|---|---|---|
| 1 | −0.006940 | −0.004567 | −0.014157 | −0.004764 | −0.014586 |
| 1.2 | −0.007036 | −0.005001 | −0.014342 | −0.005091 | −0.014909 |
| 1.3 | −0.007050 | −0.005127 | −0.014366 | −0.005178 | −0.014968 |
| 1.4 | −0.007045 | −0.005206 | −0.014354 | −0.005229 | −0.014977 |
| 1.5 | −0.007025 | −0.005250 | −0.014309 | −0.005252 | −0.014945 |
| 1.6 | −0.006991 | −0.005266 | −0.014238 | −0.005255 | −0.014881 |
| 1.7 | −0.006946 | −0.005261 | −0.014143 | −0.005239 | −0.014791 |
| 1.8 | −0.006891 | −0.005237 | −0.014029 | −0.005208 | −0.014678 |
| 2 | −0.006754 | −0.005143 | −0.013750 | −0.005106 | −0.014394 |

**Table 5:** $\Delta E_{F12}^{ep}$ (Ha) for different geminal exponents $\zeta$ (bohr$^{-1}$) using the aug-cc-pVDZ OBS and the aug-cc-pV5Z-RI ABS. The values in this table are from a separate calibration run and can differ from the values of Table 1 in the final digit (differences of less than 1 μHa).

The auxiliary basis (ABS), which is used to construct the CABS space, is the one technical choice that significantly affects the results and is the reason the ζ calibration above was performed with the aug-cc-pV5Z-RI ABS. We further investigated which ABS family to use and how large of an ABS is needed.

The OptRI family of ABSs are optimized to span the geminal RI near the electron-electron cusp and contain high angular-momentum tight basis functions but omit the valence and diffuse one-particle complement the singles relax into. Table 6 shows the consequence with the aug-cc-pVDZ OBS. For this OBS, $\Delta E_{F12}^{ep}$ is nearly independent of the ABS. However, the $\Delta E_S$ CABS corrections are not, as the OptRI ABS recovers only 23–49% of the HF basis-set error. In contrast, using the aug-cc-pV5Z orbital basis set or the RI-MP2 family of basis sets as the ABS recovers 88–97%. We note that it is possible that the OptRI+ family of basis sets[79] would perform better for the $\Delta E_S$ CABS corrections as they include additional functions designed to capture more of CABS correction.

Using a large orbital basis such as aug-cc-pV5Z as the ABS recovers the $\Delta E_S$ CABS corrections correctly, but its complement of the OBS shrinks as the OBS grows and it also lacks the high angular momentum functions the geminal RI needs to perform accurately, such that $\Delta E_{F12}^{ep}$ is in error by several percent with aug-cc-pVQZ.

Using the RI-MP2 family of basis sets as the ABS converges both $\Delta E_{F12}^{ep}$ and the $\Delta E_S$ CABS corrections. It reproduces the OptRI $\Delta E_{F12}^{ep}$ values to ~1% at every cardinality and recovers 92–93% of the $\Delta E_S$ CABS corrections obtained with the aug-cc-pV5Z orbital basis set as the ABS, corresponding to 75–90% of the HF basis-set error (Table 3). For this reason, aug-cc-pV5Z-RI is the sole ABS used for the results in Table 1.

We use a large aug-cc-pV5Z-RI ABS for every OBS rather than matching the ABS to the OBS, because otherwise the $\Delta E_S$ CABS correction can be underconverged. The $\Delta E_S$ CABS corrections grow toward the HF limit with auxiliary size at every OBS cardinal. A matched ABS undershoots the saturated value by several mHa with aug-cc-pVDZ and by $\approx 44\%$ for $FHF^-$ even with aug-cc-pVQZ. The full matched ABS, N+1 ABS, and 5Z ABS data for $\Delta E_{F12}^{ep}$ and $\Delta E_S$ for all five test molecules are shown in the Supporting Information, together with the dependence of the F12 correction on the geminal exponent with aug-cc-pVTZ (Table S3) and the validation of the electron-spin factor of Eq. 30 (Section S.III).

| ABS (aug-cc-pV5Z) | HCN $\Delta E_{F12}^{ep}$ | HCN $\Delta E_S$ CABS | $FHF^-$ $\Delta E_{F12}^{ep}$ | $FHF^-$ $\Delta E_S$ CABS |
|---|---|---|---|---|
| OptRI | −0.007004 | −0.012914 | −0.005230 | −0.015322 |
| Standard OBS | −0.007207 | −0.025331 | −0.005323 | −0.065714 |
| RI-MP2 | −0.006947 | −0.023247 | −0.005261 | −0.060868 |
| HF basis error (target) | — | −0.026265 | — | −0.067523 |

**Table 6:** $\Delta E_{F12}^{ep}$ and the electronic $\Delta E_S$ CABS correction (Ha) with aug-cc-pVDZ OBS, $\zeta = 1.7$, for the three ABS families at the aug-cc-pV5Z level. The HF basis error from Table 3 is shown for reference.

The choice of the protonic ABS has much less of an effect than the electronic ABS. Reducing the protonic ABS from a 6s5p4d3f basis set to 3s3p3d and to 2s2p changes $\Delta E_{F12}^{ep}$ by less than 0.05% for HCN (−0.006944, −0.006947, and −0.006947 Ha, respectively, with $\zeta$ = 1.7 $bohr^{-1}$ and the aug-cc-pV5Z-RI ABS), so the compact even-tempered 3s3p3d set introduced in Section III is used here. Likewise, the protonic OBS is saturated already at PB4-D. Enlarging it to PB4-F1, PB5-F, or PB6-G changes $\Delta E_{F12}^{ep}$ for HCN by at most 0.2% for aug-cc-pVDZ, and enlarging it to PB5-G changes $E_{MP2}^{ep}$ by at most 0.5% across the test set for aug-cc-pVTZ.

Using density fitting for the electronic Fock matrix in the CABS space with the aug-cc-pVQZ-jkfit auxiliary basis set changes $\Delta E_{F12}^{ep}$ by $8 \times 10^{-7}$ Ha at aug-cc-pVDZ and $7 \times 10^{-7}$ Ha at aug-cc-pVTZ, and $\Delta E_S$ by $8 \times 10^{-6}$ Ha and $4 \times 10^{-6}$ Ha, for HCN with the aug-cc-pV5Z-RI ABS. These errors are negligible, and density fitting of the electronic Fock matrix is therefore used for all calculations to avoid the in-core memory wall with larger basis sets. Using density fitting to compute the electron-proton geminal integrals does not show the same behavior. The error is negligible at the aug-cc-pVTZ level ($1 \times 10^{-7}$ Ha) but grows to $1.4 \times 10^{-4}$ Ha (5.3% of the electron-proton F12 correction) at the aug-cc-pVQZ level, because the fitting error is projected into and summed over the CABS. Hence, the geminal integrals are computed without density fitting throughout. We note that a better choice of the density-fitting RI, as in density-fitted multicomponent CC or Cholesky-decomposition DFT, might give smaller errors,[40,82] but as the memory overhead of the electron-proton geminal integrals is small for the systems in our study, we leave such exploration for future work.

Finally, we timed each step of the calculations of Table 1 on a single core of an Apple M5 Pro. Table 7 splits the wall time into multicomponent HF, the conventional multicomponent MP2 energy separated into its electronic and electron–proton parts ($E_{MP2}^{ee}$ and $E_{MP2}^{ep}$, respectively), the electron–proton F12 correction $\Delta E_{F12}^{ep}$, and the $\Delta E_S$ CABS correction.

The multicomponent MP2-F12 correction does not change how the computational cost scales with respect to system size. Like single-component MP2-F12, it adds a fixed set of intermediates, built once over the OBS and CABS, and no step scales worse than the $O(N^5)$ of the conventional MP2 correlation energy.

Computing $\Delta E_{F12}^{ep}$ as performed in this study is a significant computational expense of a multicomponent MP2-F12 calculation for smaller electronic basis sets. However, the computational expense of $\Delta E_{F12}^{ep}$ grows by only a factor of 2.4-5.0 from aug-cc-pVDZ to aug-cc-pVQZ across the set so it becomes a much smaller percentage of the computational expense for calculations with the aug-cc-pVQZ basis set. The slow growth of the computational expense of $\Delta E_{F12}^{ep}$ is because most of this expense is due to the large ABS, which is aug-cc-pV5Z-RI for every OBS. We note that, because multicomponent post-HF calculations in practice are almost always performed with QZ or larger electronic basis sets, using such a large ABS should have less of an effect on the computational expense of multicomponent F12 calculations in practice.

One final caveat should be mentioned about the timing results presented in the conventional multicomponent MP2 columns. We list them here for completeness, but they are not a fair baseline to compare against. The electron-proton MP2 step builds and transforms the full electron-proton integral tensor over the union basis instead of just the electron- or proton-pair blocks it needs, so its computation times (528 seconds for $H_5O_2^+$ at aug-cc-pVQZ) are larger than an optimized code would give. For this reason, the electron-proton MP2 timing can be larger than the electron-electron MP2 timing.

Overall, we stress that the timings should only be taken to be qualitative. Our implementation is not optimized for performance and we would expect the relative timings to become similar to single-component MP2-F12 if further optimization were to be performed.

| | OBS | HF | $E_{MP2}^{ee}$ | $E_{MP2}^{ep}$ | $\Delta E_{F12}^{ep}$ | $\Delta E_S$ | Total |
|---|---|---|---|---|---|---|---|
| HCN | DZ | 0.1 | 0.0 | 0.1 | 6.1 | 0.06 | 6.3 |
| | TZ | 0.9 | 0.1 | 0.6 | 8.9 | 0.07 | 10.6 |
| | QZ | 9.8 | 1.3 | 4.9 | 14.7 | 0.09 | 30.8 |
| $FHF^-$ | DZ | 0.1 | 0.0 | 0.1 | 6.1 | 0.06 | 6.3 |
| | TZ | 0.8 | 0.1 | 0.6 | 8.9 | 0.07 | 10.5 |
| | QZ | 6.8 | 1.4 | 5.0 | 14.7 | 0.09 | 28.0 |
| $H_5O_2^+$ | DZ | 0.4 | 0.3 | 0.4 | 31.6 | 0.21 | 32.9 |
| | TZ | 8.3 | 8.0 | 4.5 | 58.6 | 0.26 | 79.7 |
| | QZ | 306.5 | 100.0 | 527.5 | 154.4 | 0.32 | 1088.7 |
| HCCH | DZ | 0.2 | 0.0 | 0.2 | 14.2 | 0.09 | 14.7 |
| | TZ | 1.7 | 0.2 | 2.1 | 22.9 | 0.1 | 27.0 |
| | QZ | 70.9 | 23.8 | 20.4 | 39.6 | 0.13 | 154.8 |
| HNNH | DZ | 0.2 | 0.0 | 0.2 | 14.4 | 0.09 | 15.0 |
| | TZ | 2.0 | 0.2 | 2.1 | 22.0 | 0.11 | 26.4 |
| | QZ | 86.5 | 24.4 | 20.9 | 42.0 | 0.13 | 174.0 |

**Table 7:** Timing (s) of the calculations of Table 1 for the HF, $E_{MP2}^{ee}$, $E_{MP2}^{ep}$, $\Delta E_{F12}^{ep}$, and $\Delta E_S$ steps, as well as the total timing for calculations performed with the aug-cc-pVXZ electronic basis set (X=D,T,Q).

## V. Conclusions

In this study, a multicomponent MP2-F12 method was introduced to accelerate the convergence of the electron-proton correlation energy with respect to the one-particle basis set. To our knowledge, this is the first time modern F12 theory has been applied to the electron-proton correlation energy in multicomponent methods. Benchmarking calculations revealed that multicomponent MP2-F12 recovers 96-100% of the electron-proton correlation energy estimated by QZ–5Z inverse-cubic extrapolation with PB4-D held fixed for every electronic orbital basis set tested, with the aug-cc-pVDZ values already exceeding conventional multicomponent MP2 at the aug-cc-pV5Z level. The recovery percentages depend on the extrapolation, as discussed in Section IV.A. These results demonstrate the potential utility of the F12 methodology in multicomponent methods. Based on the results in this study, we think that interfacing other multicomponent methods such as multicomponent CC with modern F12 theory could be a useful direction for future research.

## Data Availability Statement

The data underlying this study are available in the published article and its Supporting Information.

## Supporting Information

Tests of the convergence of multicomponent MP2-F12 with respect to the size of the ABS, the dependence of the F12 correction on the geminal exponent at the aug-cc-pVTZ level, the coordinates of the molecules in Table 1, and the validation of the electron-spin factor of Eq. 30 (Section S.III) are given in the Supporting Information. Explicit working equations for the B intermediate are given in Section S.IV.

# Supporting Information for "An F12 Correction for Multicomponent MP2"

Mary J. Richardson,[1] Gabrielle B. Tucker,[1] and Kurt R. Brorsen[1]

Correspondence to Kurt R. Brorsen (Email: brorsenk@missouri.edu)

[1]Department of Chemistry, University of Missouri, Columbia, Missouri 65211, United States of America

## S.I Auxiliary Basis Set Convergence

The MP2-F12 results presented in Tables 1–3 use a single auxiliary basis set (ABS), aug-cc-pV5Z-RI, at all orbital-basis (OBS) cardinals. Tables S1 and S2 report the dependence of $\Delta E_{F12}^{ep}$ and of the total $\Delta E_S$ CABS corrections on this choice. We perform additional calculations with the ABS matched to the OBS cardinal and with an ABS one cardinal larger than the OBS. All remaining parameters are those of Section III ($\zeta = 1.7$ bohr⁻¹, 20-term geminal, fixed SP amplitude, relative linear-dependence threshold).

The two quantities have different auxiliary-basis requirements. $\Delta E_{F12}^{ep}$ is converged for ABS of 3Z-RI quality or larger. The matched and N+1 auxiliaries reproduce the saturated values to within 1–40 μHa with aug-cc-pVTZ and aug-cc-pVQZ. The smaller matched aug-cc-pVDZ-RI auxiliary is less complete and overestimates the correlation by up to 420 μHa (2.8 %). The $\Delta E_S$ CABS corrections converge more slowly as they increase monotonically toward the Hartree–Fock basis-set error as the ABS is enlarged. Neither the matched nor the N+1 ABS attains the saturated value at any cardinal. The difference is largest for the anion $FHF^-$, for which the matched auxiliary recovers 56% of the saturated singles at aug-cc-pVQZ. The saturated aug-cc-pV5Z-RI auxiliary is therefore used for every OBS in Tables 1–3.

| | OBS | Matched | N+1 | 5Z |
|---|---|---|---|---|
| HCN | DZ | −0.007082 | −0.007002 | −0.006947 |
| | TZ | −0.004303 | −0.004296 | −0.004304 |
| | QZ | −0.002689 | −0.002672 | −0.002672 |
| $FHF^-$ | DZ | −0.005245 | −0.005250 | −0.005261 |
| | TZ | −0.002921 | −0.002908 | −0.002920 |
| | QZ | −0.001711 | −0.001706 | −0.001706 |
| $H_5O_2^+$ | DZ | −0.005238 | −0.005249 | −0.005239 |
| | TZ | −0.002953 | −0.002933 | −0.002947 |
| | QZ | −0.001734 | −0.001731 | −0.001731 |
| HCCH | DZ | −0.014360 | −0.014264 | −0.014143 |
| | TZ | −0.008753 | −0.008754 | −0.008769 |
| | QZ | −0.005480 | −0.005445 | −0.005445 |
| HNNH | DZ | −0.015206 | −0.014855 | −0.014791 |
| | TZ | −0.009233 | −0.009170 | −0.009195 |
| | QZ | −0.005816 | −0.005798 | −0.005798 |

**Table S1:** $\Delta E_{F12}^{ep}$ computed with the ABS matched to the OBS cardinal, one cardinal larger (N+1), and at the aug-cc-pV5Z-RI level (the ABS used in the main text). Values for the two-proton systems are summed over both protons. For aug-cc-pVQZ, the N+1 ABS is aug-cc-pV5Z-RI, identical to the ABS of the main text.

| | OBS | matched | N+1 | 5Z | HF-error target |
|---|---|---|---|---|---|
| HCN | DZ | −0.020264 | −0.022148 | −0.023260 | −0.026265 |
| | TZ | −0.003969 | −0.004850 | −0.005062 | −0.006691 |
| | QZ | −0.001158 | −0.001267 | −0.001267 | −0.001216 |
| $FHF^-$ | DZ | −0.050885 | −0.055252 | −0.060869 | −0.067523 |
| | TZ | −0.008951 | −0.010706 | −0.013778 | −0.017572 |
| | QZ | −0.001743 | −0.003098 | −0.003098 | −0.003770 |
| $H_5O_2^+$ | DZ | −0.037336 | −0.042539 | −0.048257 | −0.054644 |
| | TZ | −0.007341 | −0.009304 | −0.010157 | −0.013563 |
| | QZ | −0.001876 | −0.002166 | −0.002166 | −0.002705 |
| HCCH | DZ | −0.019195 | −0.020602 | −0.021395 | −0.023810 |
| | TZ | −0.003928 | −0.004461 | −0.004675 | −0.005636 |
| | QZ | −0.001459 | −0.001559 | −0.001559 | −0.001121 |
| HNNH | DZ | −0.021409 | −0.023979 | −0.025598 | −0.028746 |
| | TZ | −0.005183 | −0.006435 | −0.006677 | −0.008341 |
| | QZ | −0.001796 | −0.001943 | −0.001943 | −0.001616 |

**Table S2:** Total $\Delta E_S$ CABS corrections (Ha) computed with the ABS matched to the OBS cardinal, set one cardinal larger (N+1), and at the aug-cc-pV5Z-RI level (the ABS used in the main text). The final column is the directly computed multicomponent Hartree–Fock basis-set error (Table 3), which the singles approach as the auxiliary is enlarged. For aug-cc-pVQZ, the N+1 ABS is aug-cc-pV5Z-RI, identical to the ABS of the main text.

Table S3 reports the dependence of $\Delta E_{F12}^{ep}$ on the geminal exponent ζ at the aug-cc-pVTZ level for HCN and $FHF^-$ with the aug-cc-pV5Z-RI ABS, complementing the aug-cc-pVDZ data of Table 5. The optimal exponent is 1.5–1.6 $bohr^{-1}$ for HCN and 2.0 $bohr^{-1}$ for $FHF^-$. The value ζ = 1.7 $bohr^{-1}$ used throughout minimizes the largest deviation from the optimum across Tables 5 and S3.

| $\zeta$ | HCN | $FHF^-$ |
|---|---|---|
| 1.5 | −0.004307 | −0.002869 |
| 1.6 | −0.004307 | −0.002898 |
| 1.7 | −0.004304 | −0.002920 |
| 1.8 | −0.004297 | −0.002936 |
| 2.0 | −0.004273 | −0.002948 |
| 2.2 | — | −0.002938 |
| 2.5 | — | −0.002900 |

**Table S3**: $\Delta E_{F12}^{ep}$ (Ha) for different geminal exponents ζ ($bohr^{-1}$) using the aug-cc-pVTZ OBS and the aug-cc-pV5Z-RI ABS.

**S.II Geometries**

All geometries were obtained from a single-component MP2/aug-cc-pVTZ optimization and held fixed across the electronic basis-set series. All coordinates are in Angstroms.

```
3
HCN
H    0.0000000000    0.0000000000    0.0135947385
C    0.0000000000    0.0000000000    1.0719978024
N    0.0000000000    0.0000000000    2.2344074590

3
FHF-
F    0.0000000000    0.0000000000   -1.1411508578
H    0.0000000000    0.0000000000    0.0000000000
F    0.0000000000    0.0000000000    1.1411508578

4
HCCH
H    0.0000000000    0.0000000000   -0.0003866377
C    0.0000000000    0.0000000000    1.0559005030
C    0.0000000000    0.0000000000    2.2640994969
H    0.0000000000    0.0000000000    3.3203866376

4
HNNH
H    1.2269474113    0.0000000000    0.5381251048
N    0.2029930400    0.0000000000    0.5904598774
N   -0.2029930400    0.0000000000   -0.5904598773
H   -1.2269474113    0.0000000000   -0.5381251048

7
H5O2+
O   -0.0448219489    0.2092183191    1.1782901541
O    0.2092183192   -0.0448219486   -1.1782901541
H    0.0340104571    0.0340104572    0.0000000000
H    0.7418838887   -0.0112096442    1.6948805385
H   -0.8255177543   -0.1268612621    1.6371943510
H   -0.0112096452    0.7418838886   -1.6948805385
H   -0.1268612610   -0.8255177544   -1.6371943510
```

## S.III Validation of the Electron-Spin Factor in Eq. 30

Equation 30 carries the prefactor $n^e = 2$. Each doubly occupied electronic orbital gives two electron-proton geminal spin channels: the proton correlates with the $\alpha$ and the $\beta$ electron at the same amplitude $t_{ep}$. The intermediates $V_{iI}$, $X_{iI}$, and $B_{iI}$ (Eqs. 14–16) are single-channel spatial integrals, so their pair sum is doubled. The electron-electron channel has no such factor as the two electrons of a closed-shell orbital form one spin-adapted pair, whereas the distinguishable proton pairs with each electron separately. We confirmed the factor two ways.

First, we built an exactly solvable model of three electronic (*i,a,a'*) and three protonic (*I,A,A'*) orbitals, with the electronic sector in the full Fock space as explicit Jordan-Wigner matrices. The Fock operators have canonical occupied blocks and nonzero virtual-CABS couplings, so the coupling terms of Eqs. 24–26 are active. The Hylleraas functional was evaluated by exact matrix algebra, minimizing the conventional doubles amplitude analytically. Two controls fix the model. At t=0, it reproduces the conventional electron-proton MP2 energy of Eq. 5 with the factor of two included. For two electrons, it reproduces the electron-electron diagonal-pair convention validated against MPQC in Section III. With both controls exact, the model correction agrees with Eq. 30, including its electron-spin factor of two. The linear and quadratic coefficients are twice those of the corresponding single-spin functional. In the expressions below, g denotes a matrix element of $r_{12}^{-1}$, so the attractive interaction has matrix element −g. This convention gives the minus signs in Eqs. S1 and S3.

We derived the same functional with spin-adapted second quantization, representing the proton by a fermionic operator that commutes with all electronic operators. Writing $g_{aA}^{iI} = \langle iI | r_{12}^{-1} | aA \rangle$ for the matrix element of the positive Coulomb kernel, $R_{xY}^{iI} = \langle iI | f_{12} | xY \rangle$ for the geminal integral of Eq. 17 restricted to the $\hat{Q}_{12}$ image blocks, the complement of the blocks in Eq. 8 ($x$ electronic, $Y$ protonic), $u_{aA}^{iI}$ for the conventional electron-proton doubles amplitude, and $\hat{T}_2$, $\hat{T}_R$, $\hat{\Phi}_{\rm ep}$ for the conventional doubles operator, the geminal operator, and the electron-proton fluctuation potential. The five matrix elements of the Hylleraas functional are then

$$\langle 0 | \hat{\Phi}_{\rm ep} \hat{T}_2 | 0 \rangle = -2 \sum_{iIaA} g_{aA}^{iI} \, u_{aA}^{iI}, \tag{S1}$$

$$\begin{aligned}\langle 0 | \hat{T}_2^{\dagger} [\hat{F}, \hat{T}_2] | 0 \rangle = {} & 2 \sum_{iIabA} F_{ab} \, u_{aA}^{iI} u_{bA}^{iI} - 2 \sum_{ijIaA} F_{ij} \, u_{aA}^{iI} u_{aA}^{jI} \\ & + 2 \sum_{iIaAB} F_{AB}^{p} \, u_{aA}^{iI} u_{aB}^{iI} - 2 \sum_{iIJaA} F_{IJ}^{p} \, u_{aA}^{iI} u_{aA}^{iJ},\end{aligned} \tag{S2}$$

$$\langle 0 | \hat{\Phi}_{\rm ep} \hat{T}_R | 0 \rangle = -2 \sum_{iIa'A} R_{a'A}^{iI} \, g_{a'A}^{iI} - 2 \sum_{iIaA'} R_{aA'}^{iI} \, g_{aA'}^{iI} - 2 \sum_{iIa'A'} R_{a'A'}^{iI} \, g_{a'A'}^{iI}, \tag{S3}$$

$$\langle 0 | \hat{T}_R^{\dagger} [\hat{F}, \hat{T}_R] | 0 \rangle = 2 \sum_{iI} \left( R^{iI} \cdot \hat{F} \cdot R^{iI} + R^{iI} \cdot \hat{F}^{p} \cdot R^{iI} - \sum_{j} F_{ij} \, R^{iI} \cdot R^{jI} - \sum_{J} F_{IJ}^{p} \, R^{iI} \cdot R^{iJ} \right), \tag{S4}$$

$$\langle 0 | \hat{T}_2^{\dagger} [\hat{F}, \hat{T}_R] | 0 \rangle = 2 \sum_{iIaA} c_{aA}^{iI} \, u_{aA}^{iI}, \tag{S5}$$

$$c_{aA}^{iI} = \sum_{a'} F_{aa'} \, R_{a'A}^{iI} + \sum_{A'} F_{AA'}^{p} \, R_{aA'}^{iI}. \tag{S6}$$

In Eq. S3, the three terms are the three $\hat{Q}_{12}$ image blocks, the complement of Eq. 8. In Eq. S4, the dot denotes contraction over the $\hat{Q}_{12}$ (virtual/CABS) slots of the two geminals, as in the $R \cdot F \cdot R$ contractions of the B intermediate (Section II.E); the first two terms place $\hat{F}$ between the excited legs of one particle and the last two are the occupied contributions, which reduce to -$2(\epsilon_i + \epsilon_I)X_{iI}$ for canonical orbitals. Eq. S6 is the *c*-intermediate of Eq. 24.

The doubles operator contains the indexed amplitudes u, while the geminal operator is multiplied by the common amplitude t. The Hylleraas functional is then given by

$$J(t,u) = t^2\langle 0|\hat{T}_R^\dagger[\hat{F},\hat{T}_R]|0\rangle + 2t\,\langle 0|\hat{T}_2^\dagger[\hat{F},\hat{T}_R]|0\rangle + \langle 0|\hat{T}_2^\dagger[\hat{F},\hat{T}_2]|0\rangle + 2t\,\langle 0|\hat{\Phi}_{\mathrm{ep}}\hat{T}_R|0\rangle + 2\,\langle 0|\hat{\Phi}_{\mathrm{ep}}\hat{T}_2|0\rangle. \quad (S7)$$

Minimizing *J* over *u* recovers Eq. 5 at t=0 and, for the t-dependent part, the assembly of Eq. 30, including its prefactor $n^e$ = 2. Eq. S1 fixes the conventional factor. Eq. S3 is the geminal V contribution and Eq. S5 is the c-intermediate coupling. Every contraction in Eqs. S1–S5 carries the same electron-spin factor of two, so it multiplies the conventional and geminal channels identically and cannot be absorbed into the cusp amplitude $t_{ep}$ (Eq. 13).

### S.IV Evaluation of the B intermediate

For one quantum proton, let **F** = **F**$_e$ + **F**$_p$, **F**$_H$ = **F** + **K**$_e$, and **D** = **F** − **QFQ**. The local Hartree potentials commute with the correlation factor, so the diagonal B intermediate satisfies

$$B_{iI} \;=\; B_{iI}^{kin} \;+\; \tfrac{1}{2}\langle\{f^2, F_H\}\rangle_{iI} \;-\; \langle f(K_e + D)f\rangle_{iI}, \tag{S8}$$

where the brackets denote a matrix element in the occupied pair $|iI\rangle$ and the kinetic contribution is given by Eq. 22. In approximation C the remaining terms are evaluated using finite-RIBS resolutions of the identity.

For real orbitals, we use *P* and *R* for electronic RIBS orbitals and *Q* and *S* for protonic RIBS orbitals, suppressing the proton label *κ*. The indices *i* and *I* denote occupied orbitals. Let *o* and *v* indicate occupied and OBS-virtual orbitals, respectively, and define

$$q_{PQ} \;=\; (1 - o_P)(1 - o_Q) - v_P v_Q, \tag{S9}$$

$$r_{PQ}^{iI} \;=\; \langle iI|f|PQ\rangle, \quad s_{PQ}^{iI} \;=\; \langle PQ|f^2|iI\rangle. \tag{S10}$$

The indicator q equals one on the virtual-CABS, CABS-virtual, and CABS-CABS blocks and zero elsewhere. The two-particle Fock and exchange matrix elements are given by

$$\mathcal{F}_{PQ,RS} \;=\; F_{PR}^{e}\delta_{QS} \;+\; \delta_{PR}F_{QS}^{p}, \tag{S11}$$

$$\mathcal{K}_{PQ,RS} \;=\; K_{PR}^{e}\delta_{QS}. \tag{S12}$$

The Hartree and completeness contributions in Eq. 16 are then

$$B_{iI}^{Hartree} \;=\; \textstyle\sum_P s_{PI}^{iI}(F^e + K^e)_{Pi} \;+\; \sum_Q s_{iQ}^{iI}F_{QI}^{p}, \tag{S13}$$

$$B_{iI}^{compl} \;=\; \textstyle\sum_{PQRS} r_{PQ}^{iI}\,[(1 - q_{PQ}q_{RS})\mathcal{F}_{PQ,RS} + \mathcal{K}_{PQ,RS}]\,r_{RS}^{iI}. \tag{S14}$$

All sums run over the finite orthonormal RIBS. We use canonical OBS orbitals and retain the occupied-CABS and OBS-virtual-CABS Fock elements. The Fock matrices include all mean-field interactions of the multicomponent HF reference, while proton exchange is absent. Expanding Eq. S14 gives the ten Fock contractions and one electronic exchange contraction used in the calculations. The Kronecker deltas reduce the two-particle expressions to one-particle matrix contractions.

The kinetic contribution is evaluated directly without a finite-RIBS closure. The squared-geminal integrals are also evaluated directly, but their contractions in Eq. S13 and the projector terms in Eq. S14 use finite-RIBS sums. The separate Gaussian fit to the squared geminal is described in Section III.